\documentclass[10pt,final,journal]{IEEEtran}

\usepackage{makeidx}
\usepackage[cmex10]{amsmath}
\usepackage{graphicx}

\newtheorem{theorem}{Theorem}
\newtheorem{proposition}{Proposition}
\newtheorem{lemma}{Lemma}
\newtheorem{remark}{Remark}
\newtheorem{definition}{Definition}
\newtheorem{example}{Example}
\newtheorem{assumption}{Assumption}

\begin{document}

\title{Optimization of Memory Usage in Tardos's Fingerprinting Codes}
\author{Koji~Nuida, Manabu~Hagiwara, Hajime~Watanabe, and~Hideki~Imai,~\IEEEmembership{Fellow,~IEEE}%
\thanks{K. Nuida, M. Hagiwara and H. Watanabe are with Research Center for Information Security (RCIS), National Institute of Advanced Industrial Science and Technology (AIST), Japan.}
\thanks{H. Imai is also with AIST-RCIS, and is with Faculty of Scientific Engineering, Chuo University, Japan.}
\thanks{This paper is the full version of the authors' work presented in: 9th Information Hiding, Saint-Malo, France, June 13, 2007.}
\thanks{This study has been sponsored by the Ministry of Economy, Trade and Industry, Japan (METI) under contract, New-generation Information Security R\&D Program, and by JSPS Grants-in-Aid for Scientific Research.}
}
\maketitle

\begin{abstract}
It is known that Tardos's collusion-secure probabilistic fingerprinting code (Tardos code; STOC'03) has length of theoretically minimal order with respect to the number of colluding users.
However, Tardos code uses certain continuous probability distribution in codeword generation, which creates some problems for practical use, in particular, it requires large extra memory.
A solution proposed so far is to use some finite probability distributions instead.
In this paper, we determine the optimal finite distribution in order to decrease extra memory amount.
By our result, the extra memory is reduced to $1/32$ of the original, or even becomes needless, in some practical setting.
Moreover, the code length is also reduced, e.g.\ to about $20.6\%$ of Tardos code asymptotically.
Finally, we address some other practical issues such as approximation errors which are inevitable in any real implementation.
\end{abstract}
\IEEEpeerreviewmaketitle
\section{Introduction}
\label{sec:introduction}

\IEEEPARstart{R}{ecent} progress in information technology has enabled us to handle easily commercial objects (such as movies, musics, customers' data) in a digital form.
This increased our convenience dramatically, however as the amount of such digital contents constantly grows, information leakage and counterfeiting, in particular those caused by authorized users, have become a serious concern.
Prevention of such illegal copying is often difficult by either technological or social reason.
An alternative solution is to embed user identification information into each content by watermarking technique, making the guilty user (called a \lq\lq pirate'') traceable from the leaked content without decreasing convenience for innocent users too much.
For this purpose, it was pointed out (\cite{BS}) that the embedded information should be designed securely against \lq\lq collusion-attacks'', that is a kind of modification of embedded information by a group of pirates.
A {\em $c$-secure code} provides such identification information which is secure against $c$ pirates or less.

It is known that Tardos's probabilistic $c$-secure code \cite{Tar} (Tardos code) has length of theoretically minimal order among all possible $c$-secure codes with respect to $c$.
The frequency of $0$s and $1$s in the codewords is decided by outputs of certain probability distribution, which is referred to as the {\em bias distribution} in this paper.
Tardos's work is a milestone in this research area because of the theoretical impact, however there are some hurdles for practical implementation, due to the property that Tardos's bias distributions are continuous.
An explicit implementation of continuous distributions would be impossible, while effects of approximation of bias distributions on the security performance have not yet been evaluated.
Moreover, (approximated values of) the outputs of the bias distribution, which should be of high accuracy to make the code $c$-secure, are supposed to be recorded throughout.
Thus large amount of extra memory is required for a practical use.

A simple solution is to replace the continuous bias distributions with {\em finite} ones.
For instance, a bias distribution with $4$ possible outputs needs only $2$ bits of memory to record one output, i.e.\ to record \lq\lq which of the four''.
This solution was first explored by Hagiwara, Hanaoka and Imai in \cite{HHI}; they established a formula of sufficient code length in terms of a given (finite) bias distribution and desired security performance.
They also proposed a \lq\lq $c$-indistinguishability'' condition for suitable bias distributions, with three concrete examples that reduce the code lengths to about $60\%$ of Tardos codes.
However, it has not yet been discussed whether their choice of bias distributions is optimal for the purpose of reducing extra memory amount.
Moreover, a problem concerning practical implementation is left unsolved as well: their code requires calculation of some \lq\lq score'' of each user, which cannot be explicitly representable in general by usual number systems on computers (e.g.\ floating-point numbers), however effects of approximation of scores have not been evaluated so far.

\begin{figure*}
\centering
\begin{picture}(0,0)%
\includegraphics{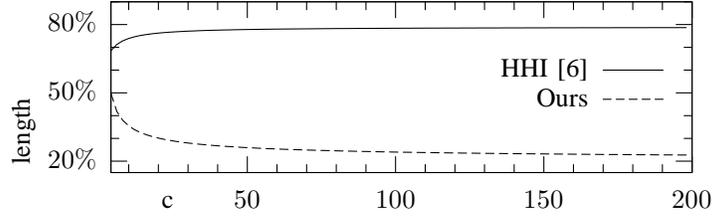}%
\end{picture}%
\begingroup
\setlength{\unitlength}{0.0200bp}%
\begin{picture}(12599,4320)(0,0)%
\put(1100,765){\makebox(0,0)[r]{\strut{}$20\%$}}%
\put(1100,2053){\makebox(0,0)[r]{\strut{}$50\%$}}%
\put(1100,3341){\makebox(0,0)[r]{\strut{}$80\%$}}%
\put(3945,0){\makebox(0,0){\strut{}$50$}}%
\put(6738,0){\makebox(0,0){\strut{}$100$}}%
\put(9532,0){\makebox(0,0){\strut{}$150$}}%
\put(12325,0){\makebox(0,0){\strut{}$200$}}%
\put(-275,1335){\rotatebox{90}{\makebox(0,0){\strut{}length}}}%
\put(2450,0){\makebox(0,0){\strut{}c}}%
\put(10374,2482){\makebox(0,0)[r]{\strut{}HHI \cite{HHI}}}%
\put(10374,1932){\makebox(0,0)[r]{\strut{}Ours}}%
\end{picture}%
\endgroup
\caption{Ratio of code lengths relative to Tardos codes}
\label{fig:length}
\end{figure*}

The aim of this paper is to solve the abovementioned problems.
First, we exhibit a strong evidence that the code lengths decrease substantially due to $c$-indistinguishability condition.
Thus, we may restrict our attention to bias distributions satisfying this condition.
Secondly, we determine the set of $c$-indistinguishable bias distributions, together with the set of the optimal ones among them (namely, those with minimal number of possible outputs).
We show that the optimal distribution has only $\lceil c/2 \rceil$ possible outputs, where $\lceil x \rceil$ denotes as usual the smallest integer $n$ with $x \leq n$; thus only $\lceil \log_2 \lceil c/2 \rceil \rceil$-bits memory are required to record one output.
(Table \ref{tab:memory} gives a numerical example, where bias distributions for Tardos codes are approximated by single-precision binary floating-point numbers.)
This shows that our result reduces the extra memory amount significantly; in particular, it even makes such extra memory needless when $c = 2$.
Moreover, we improve the code length formula in \cite{HHI} to reduce code lengths further and to evaluate effects of approximation of users' scores.
The combination of our new formula and our optimal distributions provides much shorter code lengths than Tardos codes and than \cite{HHI} (see Figure \ref{fig:length}).
We also investigate the asymptotic behavior of our code length; the ratio of our code length relative to Tardos code converges to about $20.6\%$ as $c \to \infty$.
\begin{table}[hbtp]
\centering
\caption{A comparison of required extra memory amount}
\label{tab:memory}
Case 1: $2$ pirates, $200$ users, $\mbox{error probability} \leq 10^{-11}$.\\
Case 2: $4$ pirates, $400$ users, $\mbox{error probability} \leq 10^{-11}$.\medskip\\
\begin{tabular}{|c|c|c|c|c|} \hline
\multicolumn{2}{|c|}{} & bits / position & code length & total bits \\ \hline
& Tardos & $32$ & $12\,400$ & $396\,800$ \\ \cline{2-5}
Case 1 & Ours & $\mathbf{0}$ & $\mathbf{6278}$ & $\mathbf{0}$ \\ \cline{2-5}
& \% & $\mathbf{0}$ & $\mathbf{50.6}$ & $\mathbf{0}$ \\ \hline
& Tardos & $32$ & $51\,200$ & $1\,638\,400$ \\ \cline{2-5}
Case 2 & Ours & $\mathbf{1}$ & $\mathbf{19\,750}$ & $\mathbf{19\,750}$ \\ \cline{2-5}
& \% & $\mathbf{3.1}$ & $\mathbf{38.6}$ & $\mathbf{1.2}$ \\ \hline
\end{tabular}
\end{table}

This paper is organized as follows.
After some preliminary (Section \ref{sec:summary_model}) on the model of $c$-secure codes, some preceding works, problems and notations, we observe in Section \ref{sec:c-ind} the importance of the $c$-indistinguishability condition.
Section \ref{subsec:c-ind_property} shows some properties of $c$-indistinguishable distributions; Section \ref{subsec:c-ind_determine} determines the set of $c$-indistinguishable distributions; and Section \ref{subsec:c-ind_optimal} determines the set of the optimal distributions.
Section \ref{subsec:length_formula} gives our improvement of the code length formula established in \cite{HHI}; Section \ref{subsec:length_asymptotic} investigates the asymptotic behavior of our code length; and Section \ref{subsec:numericalexample} provides some numerical examples.
We give remarks on some recent related works in Section \ref{sec:recentworks}.
Finally, Appendices are given and devoted to the proofs of some of our results.

\section{Preliminaries}
\label{sec:summary_model}

\subsection{Our Model for Collusion-Secure Codes}
\label{subsec:model}

In this subsection, we describe our model for collusion-secure codes.
In our model, a content server embeds a binary codeword $w_i = (w_{i,1},\dots,w_{i,m})$ of length $m$ into a digital content, which will be distributed to $i$-th user $u_i$, by certain watermarking technique.
{\em Pirates}, who are the adversarial users attacking the code, then make an illegal copy of the distributed content which involves a codeword possibly modified by them.
When the illegally copied content is found, the content server first extracts the embedded codeword $y = (y_1,\dots,y_m)$ (called the {\em pirated codeword}).
Some bits $y_j$ in $y$ may be broken and hence not decodable; such a bit is denoted by \lq $?$'.
Then the server executes a {\em tracing algorithm} for detecting the pirates, with the $y$ and all the $w_i$s as input.

Regarding the attack model, we assume that $\ell$ pirates try to detect the positions of (parts of) the embedded codeword from differences of their contents, and then to modify bits of the codeword in these positions by some (possibly probabilistic) algorithm, called a {\em pirates' strategy}.
This attack model is formulated as the following assumption, which was originally introduced in \cite{BS} and has been adopted in most of the preceding works (e.g.\ \cite{BS,HHI,Tar}):
\begin{assumption}[Marking Assumption]
\label{asmp:MarkingAssumption}
If all the bits $w_{i_1,j},\dots,w_{i_\ell,j}$ in codewords of the pirates $u_{i_1},\dots,u_{i_\ell}$ at the same, say $j$-th position coincide (we call such a position {\em undetectable}), then $y_j = w_{i_1,j}$.
\end{assumption}
Moreover, we also put the following assumption:
\begin{assumption}[Pirates' Knowledge]
\label{asmp:PiratesKnowledge}
Pirates have no information on the actual choice of innocent (i.e.\ non-pirate) users' codewords, other than their {\em a priori} distribution which may be publicly known.
As a result, the choice of $y$ is independent of those codewords.
\end{assumption}

To discuss the security performance of our codes, we fix the meaning of the following terms: {\em false-negative} means that the tracing algorithm outputted no pirates; {\em false-positive} means that the tracing algorithm outputted at least one innocent user; {\em tracing error} means that false-negative or false-positive (or possibly both) occurs.
A code equipped with a tracing algorithm is called {\em $c$-secure} ({\em with $\varepsilon$-error}) if the tracing error probability is bounded by a negligibly small value $\varepsilon$ provided the number of pirates is at most $c$.

\subsection{Tardos Code and Its Generalization}
\label{subsec:basicconstruction}

In this subsection we summarize the code construction and tracing algorithms of $c$-secure Tardos code \cite{Tar} and its generalization given in \cite{HHI} as follows.
First, the content server is supposed to choose the random values $0 < p^{(j)} < 1$ independently for every $1 \leq j \leq m$, according to a given probability distribution $\mathcal{P}$ which we refer to as the {\em bias distribution}.
(Details for the choices of $\mathcal{P}$ in these codes are irrelevant here and hence omitted; see the original papers for details.)
Here we only treat the bias distributions whose output values are in the open interval $(0,1)$ and which are {\em symmetric} in the following sense; we have
\begin{displaymath}
Prob((\mathcal{P}\mbox{'s output}) \in A) = Prob((\mathcal{P}\mbox{'s output}) \in 1-A)
\end{displaymath}
for any subset $A$ of the interval $(0,1)$, where $Prob$ signifies the probability and $1-A = \{1 - a \mid a \in A\}$.
(When $\mathcal{P}$ is finite, it is symmetric in this sense if and only if it outputs $a$ and $1-a$ with the same probability for any $a$.)
The resulting sequence $P = (p^{(1)},\dots,p^{(m)})$ should be stored and be kept secret throughout the scheme: due to Assumption \ref{asmp:PiratesKnowledge}, pirates may be allowed to guess the values $p^{(j)}$ from public information on $\mathcal{P}$ and the pirates' codewords, but not to know about the actual choices of $p^{(j)}$.
Then, secondly, the server chooses each codeword $w_i$ in the following probabilistic manner: $Prob(w_{i,j} = 1) = p^{(j)}$ and $Prob(w_{i,j} = 0) = 1-p^{(j)}$ for $j$-th position.
All the bits $w_{i,j}$ are supposed to be independently chosen.

In the tracing algorithm, the server calculates a score $S_i$ of each user $u_i$ by $S_i = \sum_{j=1}^{m}S_i^{(j)}$, where
\begin{displaymath}
S_i^{(j)} =
\begin{cases}
\sigma(p^{(j)}) & \mbox{if } (y_j,w_{i,j}) = (1,1)\enspace,\\
- \sigma(1-p^{(j)}) & \mbox{if } (y_j,w_{i,j}) = (1,0)\enspace,\\
0 & \mbox{if } y_j \in \{0,?\}\enspace,
\end{cases}
\end{displaymath}
with $\sigma(p) = \sqrt{ (1-p)/p }$ for $0 < p < 1$.
The output of the tracing algorithm is then the (possibly empty) list of {\em all} users $u_i$ with $S_i \geq Z$, where $Z$ is a suitably selected threshold parameter.
Details of the choices of $Z$ are also omitted here.

\subsection{Problems}
\label{subsec:problem}

A problem of Tardos code is, as we mentioned in the Introduction, that the bias distribution $\mathcal{P}$ used in his codeword generation is {\em continuous}.
An explicit implementation of such a $\mathcal{P}$ seems to be impossible.
Moreover, even if we would like to approximate this $\mathcal{P}$, e.g.\ by floating-point numbers, the original security proof does not concern effects of such inevitable approximation; and large amount of extra memory is required to record the approximated values of $\mathcal{P}$ since these values should be of high accuracy to make the code $c$-secure.

A solution proposed in \cite{HHI} is to use suitable {\em finite} bias distributions instead. They gave formulae of code length and threshold parameter corresponding to a given finite bias distribution $\mathcal{P}$, making the code $c$-secure.
Moreover, by observing the form of their formula, they also proposed a \lq\lq $c$-indistinguishability'' condition for $\mathcal{P}$ which would be effective to reduce the code lengths.

However, regarding abovementioned memory problem, the following question remained open: Is their choice of $\mathcal{P}$ optimal in terms of required memory?
Moreover, although the users' scores are irrational numbers in general, effects of approximation of scores on the tracing error probability has not yet been discussed.
In the rest of this paper, we give solutions for these problems.

\subsection{Notations}
\label{subsec:notation}

This subsection summarizes some notations used throughout this paper.
First, let the following expression
\begin{displaymath}
\{(val,prob) \mid cond\}
\end{displaymath}
signify the probability distribution such that a value $val$ is taken with probability $prob$, where $val$ and $prob$ vary subject to the condition $cond$.
Given a finite bias distribution $\mathcal{P}$, let $p_0,p_1,\dots,p_k$ denote the possible outputs in increasing order and write $q_i = Prob(\mathcal{P} \mbox{ outputs } p_i)$; thus $p_{k-i} = 1 - p_i$ and $q_{k-i} = q_i$ by the symmetry of $\mathcal{P}$.
For $1 \leq \ell \leq c$ and $0 \leq x \leq \ell$, define functions $f_{\ell,x}(p)$ and $g_{\ell,x}(p)$ for $0 < p < 1$ by
\setlength{\arraycolsep}{0.0em}
\begin{eqnarray}
f_{\ell,x}(p) &{}={}& p^x(1-p)^{\ell - x}\bigl(x\sigma(p) - (\ell-x)\sigma(1-p)\bigr) \enspace ,\nonumber\\
g_{\ell,x}(p) &{}={}& x p^{x-1}(1-p)^{\ell - x} - (\ell-x)p^x(1-p)^{\ell - x - 1} \enspace ,\nonumber
\end{eqnarray}
\setlength{\arraycolsep}{5pt}\noindent
where
\begin{displaymath}
\sigma(p) = \sqrt{ (1-p) / p } \enspace .
\end{displaymath}
These two functions satisfy the following relation $f_{\ell,x}(p) = g_{\ell,x}(p) \sqrt{p(1-p)}$.
Put
\begin{displaymath}
R_{\ell,x} = \max\{0, E_p\left[f_{\ell,x}(p)\right]\}\enspace,
\end{displaymath}
where $E_p$ signifies the expected value over outputs $p$ of $\mathcal{P}$.
Define a function $r(t)$ by 
\begin{displaymath}
r(t) = (e^t - 1 - t) / t^2 \mbox{ for } t > 0\enspace,
\end{displaymath}
and let
\begin{displaymath}
\mathcal{R}_{\ell,\mathcal{P}} 
= E_p\left[-f_{\ell,0}(p)\right] - \sum_{x = 1}^{\ell - 1}\binom{\ell}{x} R_{\ell,x} \mbox{ for } 1 \leq \ell \leq c\enspace.
\end{displaymath}
Moreover, $\log = \log_e$ denotes the natural logarithm, and $\lceil x \rceil$ denotes the smallest integer $n$ such that $n \geq x$.

\section{A Characterization of the $c$-indistinguishability}
\label{sec:c-ind}

Before solving the problems mentioned in Section \ref{subsec:problem}, we investigate the $c$-indistinguishability condition for bias distributions proposed in \cite{HHI}.
This condition was introduced for the purpose of reducing code lengths determined by the formula given in \cite{HHI}; however, it has not yet been discovered how much this condition contributes to decreasing the {\em true} tracing error probability (and hence to reducing the code length).
This section exhibits a strong evidence that this condition is in fact substantial for decreasing the error probability.

First, we recall from \cite{HHI} the following definition of the $c$-indistinguishability condition (see Section \ref{subsec:notation} for notations):
\begin{definition}
\label{defn:c-ind}
A (finite) bias distribution $\mathcal{P}$ is called {\em $c$-indistinguishable}, or {\em $c$-ind} in short, if $\sum_{x=1}^{\ell-1}\binom{\ell}{x} R_{\ell,x} = 0$ for all $2 \leq \ell \leq c$.
\end{definition}
\begin{remark}
\label{rem:c-ind}
Since the value $R_{\ell,x}$ is always nonnegative by definition, this condition is equivalent to $R_{\ell,x} = 0$ (or equivalently, $E_p\left[f_{\ell,x}(p)\right] \leq 0$) for all $2 \leq \ell \leq c$ and $1 \leq x \leq \ell-1$.
\end{remark}

Then we show that all attack strategies have the same efficiency on average {\em if and only if $\mathcal{P}$ is $c$-ind}.
This claim implies a substantial significance of the $c$-ind condition.

We start with an arbitrary finite bias distribution $\mathcal{P}$.
Let $u_1,\dots,u_\ell$ (where $\ell \leq c$) be the pirates and $w_1,\dots,w_\ell$ their codewords.
The pirates would hope none of them being outputted by the tracing algorithm, therefore they would try to create the pirated codeword $y$ so that all of their scores will be as small as possible.
For this purpose, it is necessary for the sum $S$ of their scores to be small.
By the definition of the tracing algorithm, $S$ can be decomposed as $S = S' + S''$, where $S'$ denotes the sum of pirates' bitwise scores over the undetectable positions, which is independent of the pirates' strategy due to Marking Assumption (Assumption \ref{asmp:MarkingAssumption}), and $S''$ is the sum over the remaining positions $j$ with $y_j = 1$.

Now for $1 \leq j \leq m$ and $I \subset \{1,2,\dots,\ell\}$, let $\mathcal{B}_I$ denote the event that $w_{i,j} = 1$ for $i \in I$ and $w_{i,j} = 0$ for $i \not\in I$, and let $\mathcal{B}'_I$ be the event that $\mathcal{B}_I$ occurs and $y_j = 1$.
Then the contribution of $j$-th position of their codewords for $S''$ under the event $\mathcal{B}'_I$ (where $I \neq \emptyset$ and $I \neq \{1,\dots,\ell\}$) is $x \sigma(p^{(j)}) - (\ell - x) \sigma(1-p^{(j)})$, where $x = |I|$.
Thus its expected value conditioned on $\mathcal{B}'_I$ over the choices of $p^{(j)}$ is given by
\begin{equation}
\label{eq:ex_sumofscore}
\sum_p Prob(p^{(j)} = p \mid \mathcal{B}'_I) \bigl( x \sigma(p) - (\ell - x) \sigma(1-p) \bigr) \enspace ,
\end{equation}
where the sum is taken over all possible outputs $p$ of $\mathcal{P}$.
Under this setting, our claim is expressed as the following proposition, whose proof is postponed until the end of Section \ref{subsec:c-ind_property} since it requires some results given in that section:
\begin{proposition}
\label{prop:c-ind_meaning}
The expected value (\ref{eq:ex_sumofscore}) is always $0$ if and only if $\mathcal{P}$ is $c$-indistinguishable.
\end{proposition}

Based on this observation, we restrict our attention to $c$-ind bias distributions from now on.

\section{The Optimal Bias Distribution}
\label{sec:optimaldistribution}

\subsection{Properties of the $c$-Indistinguishability Condition}
\label{subsec:c-ind_property}

In this subsection, we investigate properties of the $c$-indistinguishability condition as a preliminary for the following sections.
Proofs will be given in Appendix \ref{sec:appendix_proof_c-ind_property}.

Let $\mathcal{P}$ be a (finite) bias distribution.
First, a straightforward observation can show that $g_{\ell,\ell-x}(p) = -g_{\ell,x}(1-p)$ and $f_{\ell,\ell-x}(p) = -f_{\ell,x}(1-p)$, therefore by symmetry of $\mathcal{P}$ we have
\begin{equation}
\label{eq:relation_ex_f}
E_p \left[f_{\ell,\ell-x}(p)\right]
= -E_p \left[f_{\ell,x}(1-p)\right] 
= -E_p \left[f_{\ell,x}(p)\right] \enspace .
\end{equation}
This infers the following result concerning the case when a bias distribution becomes $c$-ind:
\begin{proposition}
\label{prop:c-ind_exiszero}
Let $\mathcal{P}$ be a (finite) bias distribution.
\begin{enumerate}
\item $\mathcal{P}$ is $c$-ind if and only if
\begin{equation}
\label{eq:c-ind_exiszero}
E_p \left[f_{\ell,x}(p)\right] = 0
\end{equation}
for any $2 \leq \ell \leq c$ and $1 \leq x \leq \ell-1$.
\item If $\ell$ is even, then (\ref{eq:c-ind_exiszero}) always holds for $x = \ell/2$.
In particular, $\mathcal{P}$ is always $2$-ind (cf.\ {\cite[Proposition 1]{HHI}}).
\item Condition (\ref{eq:c-ind_exiszero}) holds for an $\ell$ and $x = x_0$ if and only if (\ref{eq:c-ind_exiszero}) holds for this $\ell$ and $x = \ell-x_0$.
\end{enumerate}
\end{proposition}

The following recursive relations for $f_{\ell,x}$ and $g_{\ell,x}$ are key ingredients of our argument in this section:
\begin{lemma}
\label{lem:relation_f}
We have $f_{\ell-1,x}(p) = f_{\ell,x}(p) + f_{\ell,x+1}(p)$ and $g_{\ell-1,x}(p) = g_{\ell,x}(p) + g_{\ell,x+1}(p)$ for $0 \leq x \leq \ell-1$.
\end{lemma}

From this lemma, we derive the following properties.
First, the next proposition says that the $c$-ind condition simplify the value $\mathcal{R}_{\ell,\mathcal{P}}$ and makes it positive:
\begin{proposition}
\label{prop:c-ind_Requal}
If $\mathcal{P}$ is a $c$-ind distribution, then we have $\mathcal{R}_{\ell,\mathcal{P}} = E_p \left[ \sqrt{p(1-p)} \right] > 0$ for $1 \leq \ell \leq c$.
\end{proposition}

Secondly, the next lemma reduces the complexity to determine whether a given bias distribution is $c$-ind:
\begin{lemma}
\label{lem:c-ind_triangle}
If the condition (\ref{eq:c-ind_exiszero}) is satisfied for any two of the three pairs of parameters $(\ell,x) = (\ell'-1,x')$, $(\ell',x')$ and $(\ell',x'+1)$, then this condition is also satisfied for the remaining one.
\end{lemma}

Now we are able to prove the following result, which can be seen as a generalization of \cite[Proposition 1]{HHI} since any bias distribution is $1$-ind by definition:
\begin{proposition}
\label{prop:c-ind_oddtoeven}
If $c$ is odd, then any $c$-ind bias distribution is also ($c+1$)-ind.
\end{proposition}

Moreover, the following criterion of the $c$-ind condition is deduced from the above results:
\begin{proposition}
\label{prop:c-ind_criterion}
Let $c \geq 3$, and let $c'$ denote the largest odd number such that $c' \leq c$.
\begin{enumerate}
\item \label{thm_item:c-ind_criterion_bottom}
If (\ref{eq:c-ind_exiszero}) is satisfied for all parameters of the form $(\ell,x) = (c',x)$ with $1 \leq x \leq (c'-1)/2$, then $\mathcal{P}$ is $c$-ind.

\item \label{thm_item:c-ind_criterion_edge}
If $\mathcal{P}$ is $(c'-2)$-ind and (\ref{eq:c-ind_exiszero}) is satisfied for at least one parameter of the form $(c',x_0)$ with $1 \leq x_0 \leq c'-1$, then $\mathcal{P}$ is also $c$-ind.
In particular, $\mathcal{P}$ is $c$-ind if for each odd number $\ell$ with $3 \leq \ell \leq c'$, the condition (\ref{eq:c-ind_exiszero}) is satisfied for at least one parameter of the form $(\ell,x_\ell)$.
\end{enumerate}
\end{proposition}

At the end of this subsection, we give the postponed proof of Proposition \ref{prop:c-ind_meaning} in Section \ref{sec:c-ind}:
\begin{IEEEproof}
[Proof of Proposition \ref{prop:c-ind_meaning}]
First, since the value $p^{(j)}$ is assumed to be secret for the pirates, the conditional probability $Prob( y_j = 1 \mid \mathcal{B}_I \wedge (p^{(j)} = p) )$ is constant on outputs $p$ of $\mathcal{P}$, which is equal to $Prob( y_j = 1 \mid \mathcal{B}_I )$.
On the other hand, we have
\begin{displaymath}
Prob( \mathcal{B}_I \mid p^{(j)} = p ) = p^x(1-p)^{\ell - x}
\end{displaymath}
by the codeword generation.
Thus by putting
\begin{displaymath}
C = Prob( y_j = 1 \mid \mathcal{B}_I ) / Prob(\mathcal{B}'_I) \enspace ,
\end{displaymath}
we have
\setlength{\arraycolsep}{0.0em}
\begin{eqnarray}
Prob( p^{(j)} = p \mid \mathcal{B}'_I ) 
&{}={}& C \cdot Prob( (p^{(j)} = p) \wedge \mathcal{B}_I )\nonumber\\
&{}={}& C \cdot Prob( p^{(j)} = p ) p^x(1-p)^{\ell - x}\nonumber
\end{eqnarray}
\setlength{\arraycolsep}{5pt}\noindent
since $\mathcal{B}'_I = (\mathcal{B}_I \wedge (y_j = 1))$, therefore (\ref{eq:ex_sumofscore}) is equal to
\begin{displaymath}
C \sum_p Prob(p^{(j)} = p) f_{\ell,x}(p)
= C \cdot E_p \left[ f_{\ell,x}(p) \right] \enspace .
\end{displaymath}
Thus by Proposition \ref{prop:c-ind_exiszero}, this value is always $0$ regardlessly of the pirates' strategy if and only if $\mathcal{P}$ is $c$-ind.
\end{IEEEproof}

\subsection{Determining $c$-Indistinguishable Distributions}
\label{subsec:c-ind_determine}

In this subsection, we determine all the $c$-ind bias distributions $\mathcal{P}$ for every $c$, by proving in Theorem \ref{thm:SRVandQS} below that the $c$-ind bias distributions are in one-to-one correspondence with objects defined as follows:
\begin{definition}
\label{defn:QS}
We refer to a pair $\mathcal{Q} = (X,\omega)$ of a finite subset $X$ of the open interval $(-1,1)$ and a positive function $\omega > 0$ on $X$ as a {\em quadrature system}, or a {\em QS} in short, {\em of degree $d$} if we have
\begin{equation}
\label{eq:defn_QS}
\int_{-1}^{1}F(t)dt = \sum_{\xi \in X}\omega(\xi)F(\xi)
\end{equation}
for any real polynomial $F(t)$ of degree less than or equal to $d$.
We refer to the size $|X|$ of $X$ as the {\em order} of $\mathcal{Q}$, and we say that $\mathcal{Q}$ is {\em symmetric} if $-X = X$ (where $-X = \{-\xi \mid \xi \in X\}$) and $\omega(-\xi) = \omega(\xi)$ for all $\xi \in X$.
\end{definition}
\begin{example}
\label{exmp:QS}
Let $X = \{ 0, \pm \sqrt{15}/5 \}$, $\omega(0) = 8/9$ and $\omega(\pm \sqrt{15}/5) = 5/9$.
Then a direct calculation can verify that (\ref{eq:defn_QS}) holds for any $F(t)$ of degree less than or equal to $5$.
Thus $(X,\omega)$ is a symmetric QS of order $3$ and degree $5$ in the sense of Definition \ref{defn:QS}.
\end{example}

Now we give the aforementioned theorem on the one-to-one correspondence as follows, which will be proved in Appendix \ref{sec:appendix_proof_SRVandQS}:
\begin{theorem}
\label{thm:SRVandQS}
For each $c$, the $c$-ind bias distributions are in one-to-one correspondence with the symmetric QSs of degree $c-1$.
More precisely:
\begin{itemize}
\item For a symmetric QS $\mathcal{Q} = (X,\omega)$ of degree $c-1$, define a probability distribution $\mathcal{P}(\mathcal{Q})$ by
\begin{displaymath}
\mathcal{P}(\mathcal{Q}) = \left\{ \left( \frac{1 + \xi}{2}, \frac{ \omega(\xi) }{ C\sqrt{ 1-\xi^2 } } \right) \mid \xi \in X \right\}
\end{displaymath}
(see Section \ref{subsec:notation} for notation), where we put $C = \sum_{\xi \in X} \omega(\xi)/\sqrt{ 1-\xi^2 }$.
\item For a $c$-ind bias distribution $\mathcal{P} = \{(p_i,q_i) \mid 1 \leq i \leq k\}$, define a pair $\mathcal{Q}(\mathcal{P}) = (\{\xi_1,\dots,\xi_k\},\omega)$ by putting, for $1 \leq i \leq k$,
\begin{displaymath}
\xi_i = 2p_i - 1 \mbox{ and } \omega(\xi_i) = \frac{ \sqrt{p_i(1-p_i)} q_i }{ C' } \enspace ,
\end{displaymath}
where we put $C' =\sum_{i=1}^{k} \sqrt{p_i(1-p_i)} q_i/2$.
\end{itemize}
Then $\mathcal{P}(\mathcal{Q})$ is $c$-ind, $\mathcal{Q}(\mathcal{P})$ is a symmetric QS of degree $c-1$, $\mathcal{Q}(\mathcal{P}(\mathcal{Q})) = \mathcal{Q}$ and $\mathcal{P}(\mathcal{Q}(\mathcal{P})) = \mathcal{P}$.
\end{theorem}
\begin{remark}
\label{rem:eventoodd}
Note that any symmetric QS of even degree $2d$ is also a QS of degree $2d+1$ by the definition of QSs.
This fact corresponds to Proposition \ref{prop:c-ind_oddtoeven} via the one-to-one correspondence in Theorem \ref{thm:SRVandQS}.
\end{remark}

\subsection{The Optimal $c$-Indistinguishable Distribution}
\label{subsec:c-ind_optimal}

Among the $c$-ind bias distributions, in this subsection we determine the optimal ones for the purpose of reducing extra memory amount.
Owing to Proposition \ref{prop:c-ind_oddtoeven}, we may concentrate our attention on the case when $c$ is even.

First, as we mentioned in the Introduction, the optimal $c$-ind distributions are precisely the ones with minimal number of possible outputs.
By Theorem \ref{thm:SRVandQS}, such $c$-ind distributions correspond to the symmetric QSs of degree $c-1$ with minimal order; thus our task here is to determine those QSs.
However, in fact the solution of this problem has been given (in different terminology) as the following classical result:
\begin{theorem}
[e.g.\ \cite{Eng,Sze_book}]
\label{thm:GL_quadrature}
For $\nu \geq 1$, let
\begin{displaymath}
L_{\nu}(t) = \frac{1}{2^{\nu} \nu!} \left(\frac{d}{dt}\right)^{\nu}(t^2-1)^{\nu}
\end{displaymath}
be the $\nu$-th Legendre polynomial normalized as $L_{\nu}(1) = 1$.
Let $X$ be the set of zeroes of $L_{\nu}(t)$ (i.e.\ values $x$ with $L_{\nu}(x) = 0$), and put
\begin{displaymath}
\omega(\xi) = \frac{2}{ (1-\xi^2)L_{\nu}{}'(\xi)^2 } \mbox{ for } \xi \in X
\end{displaymath}
(see \cite[Section 7.3.1, p.316]{Eng} for the expression of $\omega(\xi)$).
Then $\mathcal{Q}_{\nu} = (X,\omega)$ is the unique symmetric QS of minimal order subject to the degree being $2 \nu - 1$; namely, it is a symmetric QS of order $\nu$ and degree $2\nu - 1$, while no other QS of degree $2\nu -1$ has order less than or equal to $\nu$.
\end{theorem}

For instance, $\mathcal{Q}_3$ is the QS shown in Example \ref{exmp:QS}.
We refer to the QS $\mathcal{Q}_{\nu}$ defined in this theorem as the {\em Gauss-Legendre QS}, or the {\em GL QS} in short, because of its deep relationship to the \lq\lq Gauss-Legendre quadrature formula'', that is a classical approximation method for integral (see e.g.\ \cite{Sze_book}).
Now by combining Theorems \ref{thm:SRVandQS} and \ref{thm:GL_quadrature}, we determine the optimal bias distribution (which we refer to as the {\em Gauss-Legendre distribution}, or the {\em GL distribution} in short) explicitly as follows:
\begin{theorem}
\label{thm:GLdistribution}
For $\nu \geq 1$, let
\begin{displaymath}
\widetilde{L}_{\nu}(t) = \left.\left( \frac{d}{du} \right)^{\!\nu}\!(u^2-1)^{\nu} \,\right|_{u = 2t - 1} \enspace ,
\end{displaymath}
a polynomial in $t$ of degree $\nu$.
Then the unique optimal $(2\nu)$-ind distribution $\mathcal{P} = \mathcal{P}_{2\nu}$ is given by
\begin{displaymath}
\left\{ \left(p, \frac{C}{(p(1-p))^{3/2} \widetilde{L}_{\nu}{}'(p)^2} \right) \Biggl.\Biggr|\ \widetilde{L}_{\nu}(p) = 0 \right\}
\end{displaymath}
(see Section \ref{subsec:notation} for notation), where $C$ is the normalizing constant adjusting the total probability to $1$.
This $\mathcal{P}_{2\nu}$ has $\nu$ possible outputs.
\end{theorem}

The proof of this theorem will be given in Appendix \ref{sec:appendix_proof_GLdistribution}.
Table \ref{tab:SRVsfromQS} shows the explicit GL distributions for small $c$, where the output values less than $1/2$ are omitted by symmetry.
\begin{table*}
\centering
\caption{The optimal $c$-ind distributions $\mathcal{P}_c$}
\label{tab:SRVsfromQS}
\begin{tabular}{|c||c|c|c|} \hline
\vbox to11pt{}\ $c$\, & $\widetilde{L}(t)$ & value & probability \\ \hline
$2$ & $2(2t-1)$ & $1/2$ & $1$ \\ \hline
\vbox to9pt{}$4$ & $8(6t^2-6t+1)$ & $1/2 + \sqrt{3} / 6$ & $1/2$ \\ \hline
\vbox to9pt{}$6$ & $48(2t-1)(10t^2-10t+1)$ & $1/2$ & $(20\sqrt{10} - 32) / 93$ \\
\vbox to9pt{}    && $1/2 + \sqrt{15} / 10$ & $(125 - 20\sqrt{10}) / 186$ \\ \hline
\vbox to11pt{}$8$ & $384(70t^4 - 140t^3 + 90t^2 - 20t + 1)$ & $1/2 + \sqrt{525 - 70\sqrt{30}} / 70$ & $1/4 + (41\sqrt{30} - 49\sqrt{21}) / 12$ \\
\vbox to11pt{}    && $1/2 + \sqrt{525 + 70\sqrt{30}} / 70$ & $1/4 - (41\sqrt{30} - 49\sqrt{21}) / 12$ \\ \hline
\end{tabular}
\end{table*}
\begin{remark}
\label{rem:memory_amount}
By Theorem \ref{thm:GLdistribution}, the optimal $c$-ind distribution $\mathcal{P}_c$ has $\lceil c/2 \rceil$ possible outputs, therefore only $\lceil \log_2 \lceil c/2 \rceil \rceil$-bits of memory are sufficient to record one value $p^{(j)}$ (whenever a relatively small table of possible outputs of $\mathcal{P}_c$ is held together).
As we mentioned in the Introduction, some comparison of the required memory amount between Tardos code and ours is shown in Table \ref{tab:memory} above, where we put $c=2$, $N=200$ and $\varepsilon = 10^{-11}$ in Case 1, and $c=4$, $N=400$ and $\varepsilon = 10^{-11}$ in Case 2.
Here the code lengths of our codes are calculated by using a formula given in the next section; and we assume that outputs of Tardos's continuous bias distributions are approximated by using single-precision ($4$-bytes) floating-point numbers.
The table shows that our optimal bias distributions in fact reduce the memory amount dramatically.
Note that our optimal distributions require $4$-bytes of memory or more to record one output in the (very impractical) case $c \geq 2^{32}+1 = 4\,294\,967\,297$.
However, for such $c$, the approximation of Tardos's distributions require much larger memory in order to attain comparable security.
\end{remark}

\section{Code Lengths}
\label{sec:length}

\subsection{An Improved Formula for Code Lengths}
\label{subsec:length_formula}

In this subsection, we improve the formula for code lengths and thresholds given in \cite{HHI} to reduce code lengths.
Also, we slightly modify the tracing algorithm to evaluate the effects of approximation of users' scores.
Here we do {\em not} assume that the bias distribution $\mathcal{P}$ is the optimal one determined in the previous section, since it is generally inevitable in practical implementation to perform some approximation of the optimal bias distribution (cf.\ Table \ref{tab:SRVsfromQS}).

Before stating our results, we prepare further notations (see also Section \ref{subsec:notation}).
Let $\delta \geq 0$ be a bound of approximation error of users' bitwise scores, and let $U_i$ be an approximated value of $\sigma(p_i)$ for $0 \leq i \leq k$; namely $|U_i - \sigma(p_i)| \leq \delta$.
Let $\mathcal{R}$ be a positive value such that
\begin{equation}
\label{eq:length_parameter_R}
\mathcal{R} \leq \min_{1 \leq \ell \leq c}\mathcal{R}_{\ell,\mathcal{P}} \enspace ,
\end{equation}
and let $\psi_1,\psi_2 > 0$ be approximated values of $\sigma(p_0)$ such that
\begin{equation}
\label{eq:length_parameter_sigma}
\psi_1 \leq \sigma(p_0) \leq \psi_2 \enspace .
\end{equation}
Let $\eta_1,\eta_2 > 0$ be two positive parameters, and let $x_1,x_2 > 0$ be two positive values such that
\begin{equation}
\label{eq:length_parameter_r}
x_i r(x_i) \leq \eta_i\mathcal{R}\psi_1 / c \enspace , \mbox{ for } i=1,2 \enspace .
\end{equation}
Note that $tr(t)$ is an increasing positive function for $t > 0$.
Note also that the code length given by our formula below will be reduced as the inequalities (\ref{eq:length_parameter_R}), (\ref{eq:length_parameter_sigma}) and (\ref{eq:length_parameter_r}) are getting stricter.
Moreover, choose values $A_1$ and $A_2$ so that, for $i = 1,2$,
\begin{equation}
\label{eq:length_parameter_A}
A_i \geq \frac{c}{(1-\eta_1-\eta_2/c)\mathcal{R} - 2\delta c} \cdot \frac{\psi_2}{x_i} \log \frac{1}{\varepsilon_i} \enspace ,
\end{equation}
where $\varepsilon_1$ and $\varepsilon_2$ are given security parameters related to the tracing error probability.
Note also that the code length will be decreased as the inequality (\ref{eq:length_parameter_A}) becomes stricter.

Now we define an \lq\lq approximated version'' of the tracing algorithm by the following modification:
\begin{definition}
[approximated tracing algorithm]
\label{defn:approximated_tracing}
We modify the tracing algorithm given in Section \ref{subsec:basicconstruction} as follows.
First, the approximated score $\widehat{S}_i$ of $i$-th user $u_i$ is calculated by $\widehat{S}_i = \sum_{j = 1}^{m}\widehat{S}_i^{(j)}$, where
\begin{displaymath}
\widehat{S}_i^{(j)} = 
\begin{cases}
U_{\nu_j} & \mbox{if } (y_j,w_{i,j}) = (1,1) \enspace ,\\
- U_{k-{\nu_j}} & \mbox{if } (y_j,w_{i,j}) = (1,0) \enspace ,\\
0 & \mbox{if } y_j \in \{0,?\} \enspace ,
\end{cases}
\end{displaymath}
with the index $\nu_j$ defined by $p^{(j)} = p_{\nu_j}$.
(Note that $|\widehat{S}_i - S_i| \leq m\delta$ where $S_i$ denotes the {\em true} score of $u_i$.)
Then our approximated algorithm outputs all users whose approximated score satisfies that $\widehat{S}_i \geq Z$.
\end{definition}
\begin{remark}
\label{rem:original_recovered}
Note that the original tracing algorithm is recovered when we take $U_{\nu} = \sigma(p_{\nu})$ for every $\nu$.
\end{remark}

Now sufficient code lengths and corresponding thresholds with respect to the approximated tracing algorithm are determined by the following theorem, which will be proved in Appendix \ref{sec:appendix_codelength}:
\begin{theorem}
\label{thm:approximation}
Choose the code length $m$ and the threshold $Z$ by
\setlength{\arraycolsep}{0.0em}
\begin{eqnarray}
\label{eq:newformula_length}
\hspace*{-1em}m &{}={}& A_1 + A_2 \enspace ,\\
\label{eq:newformula_threshold}
\hspace*{-1em}Z &{}={}& \left(\!\left( 1 - \frac{\eta_2}{c} \right) \frac{ \mathcal{R} }{ c } - \delta\!\right)\!A_1 + \left(\!\frac{ \eta_1 \mathcal{R} }{ c } + \delta\!\right)\!A_2
\end{eqnarray}
\setlength{\arraycolsep}{5pt}\noindent
(see above for choices of the auxiliary values), and let $N$ denote the total number of users.
Then for the approximated tracing algorithm given in Definition \ref{defn:approximated_tracing}, the false-positive probability is less than $1 - (1-\varepsilon_1)^{N-1}$ ($\leq (N-1)\varepsilon_1$); and the false-negative probability is less than $\varepsilon_2$.
Hence the total tracing error probability is bounded by $(N-1)\varepsilon_1 + \varepsilon_2$, which becomes $\varepsilon$ if we set $\varepsilon_1 = \varepsilon_2 = \varepsilon / N$.
\end{theorem}
\begin{remark}
Even if the value $R_{\ell,x}$ or $\mathcal{R}_{\ell,\mathcal{P}}$ is not explicitly representable on a computer's numeric system, all values $\mathcal{R}$, $\psi_i$ and $x_i$ can be chosen as being explicitly representable.
Moreover, $A_1$ and $A_2$, therefore the resulting code length, can be chosen from integers.
\end{remark}

Here we propose the following choice of parameters
\begin{equation}
\label{eq:choice_parameters}
(\eta_1,\eta_2) = (1/2,\sqrt{c}/2)
\end{equation}
to reduce the code length.
On the other hand, the original formula in \cite{HHI} can be recovered by putting $\delta = 0$, $\eta_1 = 1/4$ and $\eta_2 = c/2$ and by letting all of (\ref{eq:length_parameter_R}), (\ref{eq:length_parameter_sigma}), (\ref{eq:length_parameter_r}) and (\ref{eq:length_parameter_A}) be equalities.
\begin{remark}
\label{rem:obtain_figure}
Although it is somewhat complicated to compute the explicit GL distribution for large $c$, we can determine values $\mathcal{R}$, $\psi_1$ and $\psi_2$ in (\ref{eq:length_parameter_R}) and (\ref{eq:length_parameter_sigma}) by using inequalities (\ref{eq:boundofR}), (\ref{eq:bound_p0_1}) and (\ref{eq:bound_p0_2}) which will be given in Section \ref{subsec:length_asymptotic} and Appendix \ref{sec:proof_limit_p0}; thus we are still able to derive some upper bounds for the code lengths even in such cases.
Namely, if we put $\delta = 0$, then a sufficient code length $m$ making the code $c$-secure is calculated from the above values $\mathcal{R}$, $\psi_1$ and $\psi_2$ as $m = A'_1 c^2 \log(1/\varepsilon_1) + A'_2 c^2 \log(1/\varepsilon_2)$, where
\begin{displaymath}
A'_i = \frac{ \pi }{ (1-\eta_1-\eta_2/c)(c+1) x_i \tan(j_1 / \sqrt{ (c+1)^2 + a'_2 }) }
\end{displaymath}
for $i=1,2$ (see Section \ref{subsec:length_asymptotic} and Appendix \ref{sec:proof_limit_p0} for definitions of $j_1$ and $a'_2$).
By choosing security parameters $\varepsilon_1 = \varepsilon_2 = \varepsilon/N$ as in the last statement of Theorem \ref{thm:approximation}, the percentage of our code length $m$ relative to the length $100c^2 \lceil \log(N/\varepsilon) \rceil$ of Tardos code is bounded by $A'_1 + A'_2$.
Figure \ref{fig:length} above is thus obtained by plotting the values $A'_1 + A'_2$, where the lower and the upper curves correspond, respectively, to our choice (\ref{eq:choice_parameters}) of parameters and the parameters $(\eta_1,\eta_2) = (1/4,c/2)$ recovering the code length formula given in \cite{HHI}.
\end{remark}

\subsection{Asymptotic Behavior of Code Lengths}
\label{subsec:length_asymptotic}

In this subsection, we investigate properties of the GL distributions $\mathcal{P} = \mathcal{P}_c$ with $c$ even and the asymptotic behavior of the corresponding code length determined by our formula in the limit $c \to \infty$.
Proofs of results which are omitted here will be demonstrated in the Appendices.

First, we show the following bound and asymptotic behavior for the values $\mathcal{R}_{\ell,\mathcal{P}_c}$, whose proofs will be given in Appendix \ref{sec:proof_boundforR}:
\begin{proposition}
\label{prop:boundforR}
We have
\begin{equation}
\label{eq:boundofR}
\mathcal{R}_{\ell,\mathcal{P}_c} \geq \frac{ c+1 }{ c\pi } \mbox{ for all } 1 \leq \ell \leq c \enspace,
\end{equation}
and $\lim_{c \to \infty} \mathcal{R}_{\ell,\mathcal{P}_c} = 1/\pi$ for all $\ell \geq 1$.
\end{proposition}

Secondly, we are also able to show an asymptotic behavior of the value $\sigma(p_0)$.
Here we define $j_1$ to be the smallest positive zero of the $0$th-order Bessel function $J_0(t) = \sum_{i=0}^{\infty} (-1)^i (t/2)^{2i} / (i!)^2$ of the first kind; it is known that $j_1 = 2.404\,82 \cdots$.
Now the asymptotic behavior of $\sigma(p_0)$ is given as follows, which will be proved in Appendix \ref{sec:proof_limit_p0}:
\begin{proposition}
\label{prop:limit_p0}
We have $\lim_{c \to \infty} \sigma(p_0)/c = 1/j_1$.
\end{proposition}

From now, we investigate the asymptotic behavior of the code length corresponding to $\mathcal{P}_c$.
Here, for simplicity, we put $\delta = 0$, let (\ref{eq:length_parameter_R}), (\ref{eq:length_parameter_sigma}), (\ref{eq:length_parameter_r}) and (\ref{eq:length_parameter_A}) be equalities, and choose parameters $\eta_1$ and $\eta_2$ so that $\lim_{c \to \infty} \eta_1 = \eta$ with $0 < \eta < \infty$, $\lim_{c \to \infty} \eta_2 = \infty$ and $\lim_{c \to \infty} \eta_2/c = \eta'$ with $0 \leq \eta' < \infty$.
Moreover, we assume for a technical reason that $\log(1/\varepsilon_2) / \log(1/\varepsilon_1)$ does not diverge to $\infty$ when $c \to \infty$.
Then we have the following result, which will be proved in Appendix \ref{sec:proof_boundoflength_general}:
\begin{theorem}
\label{thm:boundoflength_general}
Under the above assumptions, the code length given by (\ref{eq:newformula_length}) is asymptotically
\begin{displaymath}
m \sim \frac{ \pi }{ (1-\eta-\eta') j_1 x_\infty } c^2 \log \frac{1}{\varepsilon_1} \mbox{ when } c \to \infty \enspace ,
\end{displaymath}
where $x_\infty$ is the unique positive value determined by $x_\infty r(x_\infty) = \eta / (\pi j_1)$.
\end{theorem}

By applying this theorem to our proposal (\ref{eq:choice_parameters}) of the parameters $\eta_1$ and $\eta_2$, we obtain the following result:
\begin{theorem}
\label{thm:asymptotic_length}
Put $(\eta_1,\eta_2) = (1/2,\sqrt{c}/2)$ and assume that $\log(1/\varepsilon_2) / \log(1/\varepsilon_1)$ does not diverge to $\infty$ when $c \to \infty$.
Then our code length is less than $20.6021\%$ of that of Tardos code for any sufficiently large $c$.
\end{theorem}
\begin{IEEEproof}
In this case, we have $\eta = 1/2$ and $\eta' = 0$.
By using the relations $3.141\,59 < \pi < 3.141\,60$ and $2.404\,82 < j_1 < 2.404\,83$, we have
\begin{displaymath}
x_\infty r(x_\infty) = (2\pi j_1)^{-1} > 0.066\,18 \enspace ,
\end{displaymath}
therefore it follows that $x_\infty > 0.126\,82$ (recall that $tr(t)$ is an increasing function on $t > 0$).
By these data and Theorem \ref{thm:boundoflength_general}, the percentage of our code length relative to Tardos code is asymptotically
\begin{displaymath}
\frac{\pi}{ (1 - \eta - \eta')j_1 x_\infty }
< \frac{ 3.141\,60 }{ (1/2) \cdot 2.404\,82 \cdot 0.126\,82 }
< 20.6021 \enspace .
\end{displaymath}
Thus the percentage is less than $20.6021\%$ for any sufficiently large $c$.
\end{IEEEproof}

A similar argument can be used to show that the asymptotic percentage is slightly less than $80.7028\%$ when we use the parameters $(\eta_1,\eta_2) = (1/4,c/2)$ corresponding to the formula in \cite{HHI}; in this case, we have $\eta = 1/4$, $\eta' = 1/2$, $x_\infty r(x_\infty) = (4 \pi j_1)^{-1} > 0.033\,09$ and $x_\infty > 0.064\,75$.
Thus the asymptotic behavior of our code is much better not only compared to Tardos code but also to \cite{HHI}.

\subsection{Numerical Examples}
\label{subsec:numericalexample}

Here we give some numerical examples of our code lengths and related parameters.
We use the bias distributions given in the first part of Table \ref{tab:proposal_SRV}, which approximate the GL distributions, with $c \in \{2,4,6,8\}$.
We choose approximated bitwise scores $U_i$ as in the second part of Table \ref{tab:proposal_SRV}, with approximation error $\delta = 0$ if $c = 2$ and $\delta = 10^{-5}$ if $c \in \{4,6,8\}$.
Then Table \ref{tab:codes_approximate} gives corresponding values of $\mathcal{R}$, $\psi_1$, $\psi_2$, $x_1$, $x_2$, $A_1$ and $A_2$, where we put $\varepsilon_1 = \varepsilon_2 = \varepsilon / N$, $N = 100c$, $\varepsilon = 10^{-11}$, $\eta_1 = 1/2$ and $\eta_2 = \sqrt{c}/2$.
Now by (\ref{eq:newformula_length}) and (\ref{eq:newformula_threshold}), we obtain the resulting code lengths $m$ and thresholds $Z$ as in Table \ref{tab:codes_approximate}, where the row \lq $\%$' shows percentages of our code lengths relative to Tardos codes.
On the other hand, based on results in Section \ref{subsec:length_asymptotic}, further comparison of our code lengths with those of Tardos code is given by Table \ref{tab:IMF06}, where we put $\varepsilon_1 = \varepsilon_2 = \varepsilon/N$, $N = 10^9$ and $\varepsilon = 10^{-6}$.

\begin{table}[hbtp]
\centering
\caption{Bias distributions $\mathcal{P}$ and approximated scores}
\label{tab:proposal_SRV}
\begin{tabular}{|c|c|c||c|c|c|} \hline
\ $c$\ \ & $p$ & $q$ &\ $c$\ \ & $p$ & $q$ \\ \hline
$2$ & $0.500\,00$ & $1.000\,00$
& $8$ & $0.069\,43$ & $0.248\,33$ \\ \cline{1-3}
$4$ & $0.211\,32$ & $0.500\,00$ 
& & $0.330\,01$ & $0.251\,67$ \\
& $0.788\,68$ & $0.500\,00$ 
& & $0.669\,99$ & $0.251\,67$ \\ \cline{1-3}
$6$ & $0.112\,70$ & $0.332\,01$
& & $0.930\,57$ & $0.248\,33$ \\ \cline{4-6}
& $0.500\,00$ & $0.335\,98$ & \multicolumn{3}{|c}{} \\
& $0.887\,30$ & $0.332\,01$ & \multicolumn{3}{|c}{} \\ \cline{1-3}
\end{tabular}
\quad \medskip\\
\begin{tabular}{|c||c|c|c|c|} \hline
\ $c$\ \  & $U_0$ & $U_1$ & $U_2$ & $U_3$ \\ \hline\cline{1-2}
$2$ & $1$ \\ \cline{1-3}
$4$ & $1.931\,87$ & $0.517\,63$ \\ \cline{1-4}
$6$ & $2.805\,90$ & $1$ & $0.356\,39$ \\ \cline{1-5}
$8$ & $3.661\,01$ & $1.424\,85$ & $0.701\,82$ & $0.273\,14$ \\ \hline
\end{tabular}
\end{table}
\begin{table*}[hbtp]
\centering
\caption{Auxiliary values, lengths and thresholds for the example}
\label{tab:codes_approximate}
\begin{tabular}{|cc||c|c|c|c|} \hline
 & $c$ & $2$ & $4$ & $6$ & $8$ \\ 
 & $N$ & $200$ & $400$ & $600$ & $800$ \\ \hline
Tardos & $n$ & $12\,400$ & $51\,200$ & $115\,200$ & $211\,200$ \\ \hline\hline
 & $\mathcal{R}$ & $0.5$ & $0.408$ & $0.377$ & $0.362$ \\
 & $\psi_1$ & $1$ & $1.931$ & $2.805$ & $3.661$ \\
 & $\psi_2$ & $1$ & $1.932$ & $2.806$ & $3.662$ \\
 & $x_1$ & $0.231$ & $0.184$ & $0.166$ & $0.155$ \\
 & $x_2$ & $0.315$ & $0.347$ & $0.377$ & $0.406$ \\
 & $A_1$ & $3622$ & $12\,907$ & $28\,878$ & $51\,783$ \\
 & $A_2$ & $2656$ & $6843$ & $12\,716$ & $19\,769$ \\ \hline
Ours & $n$ & $\mathbf{6278}$ & $\mathbf{19\,750}$ & $\mathbf{41\,594}$ & $\mathbf{71\,552}$ \\
 & $\%$ & $\mathbf{50.6}$ & $\mathbf{38.6}$ & $\mathbf{36.1}$ & $\mathbf{33.9}$ \\
 & $Z$ & $917.3\cdots$ & $1336.317\,86$ & $1843.450\,24\cdots$ & $2375.914\,48\cdots$ \\ \hline
\end{tabular}
\end{table*}
\begin{table*}
\centering
\caption{Another comparison of code lengths for $N = 10^9$ and $\varepsilon = 10^{-6}$}
\label{tab:IMF06}
\begin{tabular}{|c||c|c|c|c|c||c|} \hline
$c$ & $4$ & $8$ & $16$ & $32$ & $64$ & $\to \infty$ \\ \hline
\vbox to9pt{}Tardos & $5.60 \times 10^4$ & $2.24 \times 10^5$ & $8.96 \times 10^5$ & $3.58 \times 10^6$ & $1.43 \times 10^7$ & $100\%$ \\
\vbox to9pt{}Ours & $2.18 \times 10^4$ & $7.72 \times 10^4$ & $2.78 \times 10^5$ & $1.01 \times 10^6$ & $3.75 \times 10^6$ & $20.6\%$ \\ \hline
\end{tabular}
\end{table*}

These examples show that our result in this paper indeed reduces the code lengths.

\section{Remarks on Recent Related Works}
\label{sec:recentworks}

At the time when the preliminary version of this paper was written, our code lengths given in Section \ref{sec:length} were to our best knowledge the shortest among known $c$-secure codes (at least for $c \geq 4$).
After that, some recent works \cite{IM06,IM07,SKC,SVCT} on Tardos code have succeeded to reduce the code lengths, by strictly improving the evaluation of tracing error probabilities and slightly modifying some parameters or even the tracing algorithm itself; their new code lengths are in fact shorter than ours.
However, in their works, the problems such as large memory amount and impossibility of explicit implementation, mentioned and solved in this paper, are not concerned.
For instance, their schemes still use continuous bias distributions but they did not show suitable ways to implement or approximate their continuous distributions for practical use.\\
\indent
Therefore, these recent results do not completely supersede the work in this paper, in particular, the most significant part regarding reduction of extra memory amount.
In fact, these results show that there remains a room for reducing the length of our code.
Indeed, we would like to announce that our recent successive study has achieved code lengths even shorter than the abovementioned works, by using (approximation of) the GL distributions, improving our tracing algorithm, and tightly evaluating its tracing error probability.
The details of the successive result will be presented in a forthcoming paper.

\section{Conclusion}
\label{sec:conclusion}

We have discussed the problems of Tardos's fingerprinting code \cite{Tar} regarding its practical use, such as large required memory and impossibility of explicit implementation, mainly due to continuity of probability distributions used in its codeword generation.
We investigated the finite probability distributions used in the preceding improvement \cite{HHI} of Tardos code, and determined the optimal distributions for the purpose of reducing memory amount.
As Table \ref{tab:memory} shows, the memory amount is indeed reduced dramatically by our result.
We also reduced the code lengths significantly by improving the formula of code lengths given in \cite{HHI}; and evaluated effects of approximation on security performance of our codes in a practical setting.

\section*{Acknowledgment}
The authors would like to express their gratitude to Kazuto Ogawa and Satoshi Fujitsu at Japan Broadcasting Corporation (NHK), and to Takashi Kitagawa, Rui Zhang and Kirill Morozov at National Institute of Advanced Industrial Science and Technology (AIST), for several significant comments.

\appendices

\section{Proofs of Results in Section \ref{subsec:c-ind_property}}
\label{sec:appendix_proof_c-ind_property}

Here we give the proofs of our results in Section \ref{subsec:c-ind_property}.
\begin{IEEEproof}
[Proof of Proposition \ref{prop:c-ind_exiszero}]
First, the following property is easily derived from (\ref{eq:relation_ex_f}): $E_p \left[f_{\ell,\ell-x}(p)\right] \geq 0$ for all $2 \leq \ell \leq c$ and $1 \leq x \leq \ell-1$ if and only if $E_p \left[f_{\ell,x}(p)\right] \leq 0$ for all $2 \leq \ell \leq c$ and $1 \leq x \leq \ell-1$.
Thus the first claim follows from Remark \ref{rem:c-ind}.
The other claims are also straightforward by (\ref{eq:relation_ex_f}).
\end{IEEEproof}
\begin{IEEEproof}
[Proof of Lemma \ref{lem:relation_f}]
First, an elementary analysis shows that
\begin{equation}
\label{eq:expressionofg}
g_{\ell,x}(p) = \frac{d}{dp} \left( p^x(1-p)^{\ell-x} \right) \enspace ,
\end{equation}
therefore the second claim follows from the equality
\begin{displaymath}
p^x(1-p)^{\ell-x} + p^{x+1}(1-p)^{\ell-x-1} = p^x(1-p)^{\ell-1-x} \enspace .
\end{displaymath}
Now the first claim is also derived from the relation $f_{\ell,x}(p) = g_{\ell,x}(p) \sqrt{p(1-p)}$.
\end{IEEEproof}
\begin{IEEEproof}
[Proof of Proposition \ref{prop:c-ind_Requal}]
By the assumption on $\mathcal{P}$, we have $\mathcal{R}_{\ell,\mathcal{P}} = E_p \left[-f_{\ell,0}(p)\right]$ for any $1 \leq \ell \leq c$.
Thus Lemma \ref{lem:relation_f} infers that $\mathcal{R}_{\ell, \mathcal{P}} - \mathcal{R}_{\ell-1, \mathcal{P}} = E_p \left[f_{\ell,1}(p)\right]$ for any $2 \leq \ell \leq c$, therefore we have $\mathcal{R}_{\ell, \mathcal{P}} - \mathcal{R}_{\ell-1, \mathcal{P}} = 0$ by Proposition \ref{prop:c-ind_exiszero}.
Hence $\mathcal{R}_{\ell,\mathcal{P}} = \mathcal{R}_{1,\mathcal{P}} = E_p \left[ \sqrt{p(1-p)} \right]$, as desired.
\end{IEEEproof}
\begin{IEEEproof}
[Proof of Lemma \ref{lem:c-ind_triangle}]
We have $E_p \left[f_{\ell'-1,x'}(p)\right] = E_p \left[f_{\ell',x'}(p)\right] + E_p \left[f_{\ell',x'+1}(p)\right]$ by Lemma \ref{lem:relation_f}; thus all of the three terms become zero whenever any two of them are.
\end{IEEEproof}
\begin{IEEEproof}
[Proof of Proposition \ref{prop:c-ind_oddtoeven}]
Since $\mathcal{P}$ is $c$-ind, Proposition \ref{prop:c-ind_exiszero} infers that (\ref{eq:c-ind_exiszero}) is satisfied for all parameters of the form $(\ell,x)$ with $2 \leq \ell \leq c$ and for $(c+1,x_0)$, where $x_0 = (c+1)/2$.
Thus by Lemma \ref{lem:c-ind_triangle} and induction on $\nu$, it follows for all $\nu$ that (\ref{eq:c-ind_exiszero}) is satisfied for parameters of the form $(c+1,x_0 \pm \nu)$.
Hence $\mathcal{P}$ is $(c+1)$-ind by Claim 1 of Proposition \ref{prop:c-ind_exiszero}.
\end{IEEEproof}
\begin{IEEEproof}
[Proof of Proposition \ref{prop:c-ind_criterion}]
By Proposition \ref{prop:c-ind_oddtoeven}, it suffices for both of the two claims to prove that $\mathcal{P}$ is $c'$-ind.

First, we argue the claim 1.
By the assumption and Claim 3 of Proposition \ref{prop:c-ind_exiszero}, the condition (\ref{eq:c-ind_exiszero}) is satisfied for all parameters $(c',x)$ with $1 \leq x \leq c'-1$, therefore Lemma \ref{lem:c-ind_triangle} infers that it is also satisfied for all parameters $(c'-1,x)$ with $1 \leq x \leq c'-2$.
Similarly, it is inductively derived that (\ref{eq:c-ind_exiszero}) is satisfied for all parameters $(\ell,x)$ with $2 \leq \ell \leq c'$ and $1 \leq \ell \leq \ell-1$.
Thus $\mathcal{P}$ is $c'$-ind by Claim 1 of Proposition \ref{prop:c-ind_exiszero}.

Secondly, we prove the claim 2.
The assumption and Proposition \ref{prop:c-ind_oddtoeven} infer that $\mathcal{P}$ is $(c'-1)$-ind, thus (\ref{eq:c-ind_exiszero}) is satisfied for all parameters $(c'-1,x)$ with $1 \leq x \leq c'-2$.
Since (\ref{eq:c-ind_exiszero}) is satisfied for the parameter $(c',x_0)$ in the statement, the same argument as Proposition \ref{prop:c-ind_oddtoeven} shows that (\ref{eq:c-ind_exiszero}) is also satisfied for all parameters $(c',x)$ with $1 \leq x \leq c'-1$.
Hence $\mathcal{P}$ is $c'$-ind by Claim 1 of Proposition \ref{prop:c-ind_exiszero}.
\end{IEEEproof}

\section{Proof of Theorem \ref{thm:SRVandQS}}
\label{sec:appendix_proof_SRVandQS}

Here we give the proof of Theorem \ref{thm:SRVandQS}.
First, we show that $\mathcal{P} = \mathcal{P}(\mathcal{Q})$ is a $c$-ind bias distribution for any symmetric QS $\mathcal{Q}$ of degree $c-1$.
A straightforward calculation can show that this $\mathcal{P}$ is indeed a finite probability distribution; the outputs of $\mathcal{P}$ lie in the interval $(0,1)$ since $X$ is a subset of the interval $(-1,1)$; and $\mathcal{P}$ is symmetric since $\mathcal{Q}$ is symmetric.
Thus the remaining task is, by Claim 2 of Proposition \ref{prop:c-ind_criterion}, to show that $E_p \left[f_{\ell,1}(p)\right] = 0$ for all $2 \leq \ell \leq c$.
Now recall the relation $f_{\ell,1}(p) = \sqrt{p(1-p)} g_{\ell,1}(p)$.
Since $g_{\ell,1}$ is a polynomial of degree $\ell-1$ ($\leq c-1$), we have
\setlength{\arraycolsep}{0.0em}
\begin{eqnarray}
&&E_p \left[ \sqrt{p(1-p)} g_{\ell,1}(p) \right]\nonumber\\
&&= \sum_{\xi \in X} \sqrt{ \frac{1+\xi}{2} \cdot \frac{1-\xi}{2} }\ g_{\ell,1}\!\left( \frac{1+\xi}{2} \right) \cdot \frac{ \omega(\xi) }{ C \sqrt{ 1-\xi^2 } }\nonumber\\
&&= \frac{1}{2C} \sum_{\xi \in X} \omega(\xi)g_{\ell,1}\!\left( \frac{ 1+\xi }{ 2 } \right)\nonumber\\
&&= \frac{1}{2C} \int_{-1}^{1} g_{\ell,1}\!\left( \frac{ 1+t }{ 2 } \right)dt\label{eq:SRVandQS_1_1}\\
&&= \frac{1}{C} \int_{0}^{1} g_{\ell,1}(z)\,dz\nonumber\\
&&= 0\label{eq:SRVandQS_1_2}
\end{eqnarray}
\setlength{\arraycolsep}{5pt}\noindent
(here (\ref{eq:SRVandQS_1_1}) follows since $\mathcal{Q}$ is a QS of degree $c-1$, while (\ref{eq:SRVandQS_1_2}) is derived from (\ref{eq:expressionofg})).
Thus $\mathcal{P}(\mathcal{Q})$ is $c$-ind.

Secondly, we show that $\mathcal{Q} = \mathcal{Q}(\mathcal{P})$ is a symmetric QS of degree $c-1$ for any $c$-ind distribution $\mathcal{P}$.
The set $X$ is included in the interval $(-1,1)$ since $0 < p_i < 1$ for all $i$, while $\mathcal{Q}$ is symmetric since $\mathcal{P}$ is symmetric.
Thus the remaining task is to show that $\int_{-1}^{1}F(t)dt = \sum_{\xi \in X}\omega(\xi)F(\xi)$ for any polynomial $F(t)$ of degree less than or equal to $c-1$.
Now observe that any such $F(t)$ can be expressed as a linear combination of the polynomials $g_{\ell,1}(\frac{1+t}{2})$ of degree $\ell-1$ for $2 \leq \ell \leq c$ and a constant polynomial $1$, while $\sum_i \omega(\xi_i) = 2 = \int_{-1}^{1}1\,dt$ by definition.
Thus it suffices to show the above claim only for $F(t) = g_{\ell,1}(\frac{1+t}{2})$ with $2 \leq \ell \leq c$.
For this claim, we have
\setlength{\arraycolsep}{0.0em}
\begin{eqnarray}
C'\! \int_{-1}^{1} g_{\ell,1}\!\left( \frac{ 1+t }{ 2 } \right)dt
&{}={}& 2C'\! \int_{0}^{1} g_{\ell,1}(z)\,dz\nonumber\\
&{}={}& 0\label{eq:SRVandQS_2_1}\\
&{}={}& E_p\left[ \sqrt{p(1-p)} g_{\ell,1}(p) \right]\label{eq:SRVandQS_2_2}\\
&{}={}& C' \sum_{i=1}^{k} \omega(\xi_i) g_{\ell,1}\!\left(\frac{ 1+\xi_i }{ 2 } \right)\nonumber
\end{eqnarray}
\setlength{\arraycolsep}{5pt}\noindent
(here (\ref{eq:SRVandQS_2_1}) is derived from (\ref{eq:expressionofg}), while (\ref{eq:SRVandQS_2_2}) follows from Claim 1 of Proposition \ref{prop:c-ind_exiszero}).
Thus $\mathcal{Q}(\mathcal{P})$ is a symmetric QS of degree $c-1$.

Finally, since $\sum_{\xi \in X} \omega(\xi) = 2$ and $\sum_i q_i = 1$, a straightforward computation can verify the relations $\mathcal{Q}(\mathcal{P}(\mathcal{Q})) = \mathcal{Q}$ and $\mathcal{P}(\mathcal{Q}(\mathcal{P})) = \mathcal{P}$.
Hence the proof of Theorem \ref{thm:SRVandQS} is concluded.

\section{Proof of Theorem \ref{thm:GLdistribution}}
\label{sec:appendix_proof_GLdistribution}

Here we give the proof of Theorem \ref{thm:GLdistribution}.
Put
\begin{displaymath}
\widehat{L}_{\nu}(t) = L_{\nu}(2t-1) \enspace ,
\end{displaymath}
which is proportional to $\widetilde{L}_{\nu}$.
First, note that $\widehat{L}_{\nu}(\frac{1+\xi}{2}) = 0$ if and only if $L_{\nu}(\xi) = 0$, thus the set of outputs of $\mathcal{P}_c = \mathcal{P}(\mathcal{Q}_{\nu})$ with $c = 2\nu$ is (by definition) the set of zeroes of $\widehat{L}_{\nu}$, which coincides with the set of zeroes of $\widetilde{L}_{\nu}$ and consists of $\nu$ elements (see Theorem \ref{thm:GL_quadrature}).
Now note that $1-\xi^2 = 4p(1-p)$ if $p = (1+\xi)/2$, while
\setlength{\arraycolsep}{0.0em}
\begin{eqnarray}
\left.\frac{d}{dt}L_{\nu}(t)\right|_{t = 2p-1}
&{}={}& \left.\frac{d}{dt}\widehat{L}_{\nu}\!\left( \frac{1+t}{2} \right) \right|_{t = 2p-1}\nonumber\\
&{}={}& \left.\frac{1}{2}\left( \left.\frac{d}{du}\widehat{L}_{\nu}(u)\right|_{u = (1+t)/2} \right)\right|_{t = 2p-1}\nonumber\\
&{}={}& \frac{1}{2} \left.\frac{d}{du} \widehat{L}_{\nu}(u)\right|_{u = p}
= C'' \widetilde{L}_{\nu}{}'(p) \enspace ,\nonumber
\end{eqnarray}
\setlength{\arraycolsep}{5pt}\noindent
where $C''$ is some constant.
Thus the probability of $\mathcal{P}_c$ taking the value $p = (1+\xi)/2$ with $\xi \in X$ is
\setlength{\arraycolsep}{0.0em}
\begin{eqnarray}
\frac{\omega(\xi)}{C \sqrt{1-\xi^2}}
&{}={}& \frac{2}{C (1-\xi^2)^{3/2} L_{\nu}{}'(\xi)^2}\nonumber\\
&{}={}& \frac{1}{4CC''{}^2 \bigl(p(1-p)\bigr)^{3/2} \widetilde{L}_{\nu}{}'(p)^2} \enspace .\nonumber
\end{eqnarray}
\setlength{\arraycolsep}{5pt}\noindent
Hence the claim follows, since the factor $1/(4CC''{}^2)$ above is common for all $p$, concluding the proof.

\section{Proof of Theorem \ref{thm:approximation}}
\label{sec:appendix_codelength}

Here we give the proof of Theorem \ref{thm:approximation} by evaluating the probabilities of false-negative and of false-positive.
This will be done by basically the same argument as \cite{HHI} except for some slight modifications.

In what follows, let $\hat{x}_1$ and $\hat{x}_2$ be two positive parameters, and put $\alpha = \hat{x}_1 / \sigma(p_0)$ and $\beta = \hat{x}_2 / (c\sigma(p_0))$.
Before giving our proof, we recall the following fundamental tool in probability theory which is used in our argument (as well as in \cite{HHI}):
\begin{lemma}
[Markov's Inequality]
\label{lem:MarkovInequality}
Let $Y$ be a finite positive random variable and $t > 0$.
Then we have
\begin{displaymath}
Prob(Y > t) < \frac{E\left[Y\right]}{t} \mbox{ and } Prob(Y \geq t) \leq \frac{E\left[Y\right]}{t} \enspace ,
\end{displaymath}
where $E\left[Y\right]$ denotes the expected value of $Y$.
\end{lemma}

Now we give the following proposition, which is a slight modification of \cite[Lemma 1]{HHI} and which concerns the false-positive probability of our code:
\begin{proposition}
[cf.\ {\cite[Lemma 1]{HHI}}]
\label{prop:HHI_innocent}
Let $u_i$ be an innocent user.
For any fixed $P = (p^{(1)},\dots,p^{(m)})$, any fixed $y = (y_1,\dots,y_m)$ and any $t > 0$, we have
\begin{equation}
\label{eq:HHI_innocent}
Prob(S_i \geq t) < e^{r(\hat{x}_1)\alpha^2 m - \alpha t} \enspace ,
\end{equation}
where the probability is taken over the codewords of $u_i$ chosen according to the $P$.
\end{proposition}
\begin{IEEEproof}
The proof is almost the same as that of \cite[Lemma 1]{HHI}, except for some differences explained below.
First, \cite[Lemma 1]{HHI} showed an inequality similar to (\ref{eq:HHI_innocent}) for the probability $Prob(S_i > Z)$ under the assumption that $\alpha \sigma(p_0) < \hat{x}_1$ (note that $\hat{x}_1$ is simply denoted by $x_1$ in \cite{HHI}); however, the same proof is actually able to prove the same inequality for a slightly larger probability $Prob(S_i \geq Z)$ under the weaker assumption that $\alpha \sigma(p_0) \leq \hat{x}_1$. 
This follows from the observation that the bound $1 + r_1 \alpha^2 \leq e^{r_1 \alpha ^ 2}$ used in the original proof is indeed a {\em strict} inequality.
Secondly, \cite[Lemma 1]{HHI} was originally proved only when $Z$ is the threshold, however a careful reading of the proof can reveal that the property of $Z$ {\em being the threshold} is not used in there; therefore that proof is still valid even if $Z$ is just an {\em arbitrary} positive parameter.
Now our claim follows by combining these two observations.
\end{IEEEproof}
\begin{remark}
\label{rem:HHI_proof}
The proof of {\cite[Lemma 1]{HHI}} is still valid (so is that of the above proposition) even if $y$ is an {\em arbitrary} codeword with $y_j \in \{0,1,?\}$ {\em which need not satisfy the Marking Assumption}, only the required property of $y$ is that it is independent of the codeword of $u_i$ (see Assumption \ref{asmp:PiratesKnowledge}).
\end{remark}

On the other hand, the next proposition, which is a slight modification of \cite[Lemma 2]{HHI}, concerns the false-negative probability:
\begin{proposition}
[cf.\ {\cite[Lemma 2]{HHI}}]
\label{prop:HHI_guilty}
Let $u_1,\dots,u_\ell$ be the pirates with $\ell \leq c$, and $t > 0$.
Then for any fixed pirates' strategy, we have
\begin{equation}
E \left[e^{-\beta \sum_{i = 1}^{\ell} S_i}\right] \leq e^{ \beta (c\beta r(\hat{x}_2) - \mathcal{R}_{\ell,\mathcal{P}}) m } \enspace ,
\end{equation}
where the expected value is taken over all $P$, all codewords of pirates and all $y$, which are chosen according to $\mathcal{P}$, $P$ and the pirates' strategy, respectively.
Hence by Markov's Inequality, we have
\setlength{\arraycolsep}{0.0em}
\begin{eqnarray}
Prob(S_i < t \mbox{ for all } i) 
&{}\leq{}& Prob \left(\sum_{i=1}^{\ell} S_i < \ell t\right) \nonumber\\
&{}\leq{}& Prob \left( e^{-\beta \sum_i S_i} > e^{-\beta \ell t} \right) \nonumber\\
&{}<{}& E \left[e^{-\beta \sum_i S_i}\right] / e^{-\beta \ell t} \nonumber\\
&{}\leq{}& e^{ \beta (c\beta r(\hat{x}_2) - \mathcal{R}_{\ell,\mathcal{P}}) m + \beta \ell t } \enspace .\nonumber
\end{eqnarray}
\setlength{\arraycolsep}{5pt}
\end{proposition}
\begin{IEEEproof}
The proof is basically the same as \cite[Lemma 2]{HHI}; it also works in our situation by noticing the following points only.
First, the original proof allows the pirates' strategy to be probabilistic, though it was not clarified.
Secondly, although \cite{HHI} only considers the restricted case that $y$ contains no bit \lq $?$', an argument appeared in {\cite{Tar}} can generalize the proof in \cite{HHI} to our situation where $y$ may contain \lq $?$'.
\end{IEEEproof}

Now we start to prove Theorem \ref{thm:approximation}.
First, recall that in the approximated tracing algorithm given in Definition \ref{defn:approximated_tracing}, a user is outputted if and only if $\widehat{S} \geq Z$ where $\widehat{S}$ denotes the approximated score.
Now $\widehat{S} \geq Z$ infers $S \geq Z - m\delta$ and $\widehat{S} < Z$ infers $S < Z + m\delta$ by definition of $\delta$; thus to achieve $Prob(u_{i_0} \mbox{ is outputted}) < \varepsilon_1$, where $u_{i_0}$ is an arbitrarily fixed innocent user, and to achieve $Prob(\mbox{no pirate is outputted}) < \varepsilon_2$ as well, it suffices to satisfy the following two bounds
\setlength{\arraycolsep}{0.0em}
\begin{eqnarray}
\label{eq:securitywithapproximation_1_1}
Prob(S \geq Z - m\delta) &{}<{}& \varepsilon_1 \enspace ,\\
\label{eq:securitywithapproximation_1_2}
Prob(S_i < Z + m\delta \mbox{ for all } i) &{}<{}& \varepsilon_2 \enspace ,
\end{eqnarray}
\setlength{\arraycolsep}{5pt}\noindent
where $S$ denotes an arbitrarily fixed innocent user's true score and $S_1,\dots,S_\ell$ (with $\ell \leq c$) denote the $\ell$ pirates' true scores.
Now by Propositions \ref{prop:HHI_innocent} and \ref{prop:HHI_guilty}, the following conditions yield (\ref{eq:securitywithapproximation_1_1}) and (\ref{eq:securitywithapproximation_1_2}):
\setlength{\arraycolsep}{0.0em}
\begin{eqnarray}
r(\hat{x}_1)\alpha^2 m - \alpha (Z - m\delta) &{}\leq{}& \log \varepsilon_1 \enspace, \nonumber\\
\beta( c\beta r(\hat{x}_2) - \mathcal{R}_{\ell,\mathcal{P}} ) m + \beta \ell( Z + m\delta ) &{}\leq{}& \log \varepsilon_2 \enspace .\nonumber
\end{eqnarray}
\setlength{\arraycolsep}{5pt}\noindent
Moreover, since the values $m$, $Z$, $\beta$ and $\delta$ are all nonnegative, the following conditions
\setlength{\arraycolsep}{0.0em}
\begin{eqnarray}
\label{eq:securitywithapproximation_2_1}
r(\hat{x}_1)\alpha^2 m - \alpha (Z - m\delta) &{}={}& \log \varepsilon_1 \enspace ,\\
\label{eq:securitywithapproximation_2_2}
\beta( c\beta r(\hat{x}_2) - \mathcal{R} ) m + \beta c( Z + m\delta ) &{}={}& \log \varepsilon_2
\end{eqnarray}
\setlength{\arraycolsep}{5pt}\noindent
also yield (\ref{eq:securitywithapproximation_1_1}) and (\ref{eq:securitywithapproximation_1_2}).
Now if we solve equations (\ref{eq:securitywithapproximation_2_1}) and (\ref{eq:securitywithapproximation_2_2}) in $m$ and $Z$, where $\hat{x}_1$ and $\hat{x}_2$ are determined by $\hat{x}_i r(\hat{x}_i) = \sigma(p_0) \eta_i \mathcal{R} / c$ for $i=1,2$ and $\alpha$ and $\beta$ are determined as above, then the code length $m$ and the threshold $Z$ are given by
\begin{equation}
\label{eq:newlength_precise_1}
m = \frac{ c \sigma(p_0) }{ C } \left( \frac{1}{\hat{x}_1} \log \frac{1}{\varepsilon_1} + \frac{1}{\hat{x}_2} \log\frac{1}{\varepsilon_2} \right)
\end{equation}
(where $C = (1 - \eta_1 - \eta_2 / c) \mathcal{R} - 2 \delta c$) and
\begin{equation}
\label{eq:newlength_precise_2}
Z = \frac{ \sigma(p_0) }{ C } \left( \frac{ (1 - \eta_2/c) \mathcal{R} - \delta c }{\hat{x}_1} \log\frac{1}{\varepsilon_1} + \frac{ \eta_1 \mathcal{R} + \delta c }{\hat{x}_2} \log\frac{1}{\varepsilon_2} \right)
\end{equation}
which generalize the formula in \cite{HHI} (the original is recovered by putting $\eta_1 = 1/4$, $\eta_2 = c/2$ and $\delta = 0$).
Moreover, if we take the values $\psi_i$, $x_i$ and $A_i$ ($i=1,2$) as in Section \ref{subsec:length_formula}, and determine the modified code length $\widehat{m}$ and threshold $\widehat{Z}$ by
\setlength{\arraycolsep}{0.0em}
\begin{eqnarray}
\label{eq:newlength_approx_1}
\hspace*{-1.5em}\widehat{m} &{}={}& A_1 + A_2 \enspace ,\\
\label{eq:newlength_approx_2}
\hspace*{-1.5em}\widehat{Z} &{}={}& \left( \left( 1 - \frac{\eta_2}{c} \right) \frac{ \mathcal{R} }{ c } - \delta \right) A_1 + \left( \frac{ \eta_1 \mathcal{R} }{ c } + \delta \right) A_2 \enspace ,
\end{eqnarray}
\setlength{\arraycolsep}{5pt}\noindent
then by comparing the pair of (\ref{eq:newlength_precise_1}) and (\ref{eq:newlength_precise_2}) with the pair of (\ref{eq:newlength_approx_1}) and (\ref{eq:newlength_approx_2}), we can show that conditions (\ref{eq:securitywithapproximation_1_1}) and (\ref{eq:securitywithapproximation_1_2}) are satisfied with $\varepsilon_i$ replaced by $e^{-k_i}$ where
\begin{displaymath}
k_i = \frac{ C }{ c\sigma(p_0) } x_i A_i \mbox{ for } i=1,2 \enspace .
\end{displaymath}
Since $\psi_2 \geq \sigma(p_0)$ and $x_i \leq \hat{x}_i$ ($i=1,2$) by definition, we have $e^{-k_i} \leq \varepsilon_i$ by the choice of $A_i$.
Thus the code length $\widehat{m}$ and threshold $\widehat{Z}$, which are precisely those chosen in Theorem \ref{thm:approximation} (see (\ref{eq:newformula_length}) and (\ref{eq:newformula_threshold})), provide the desired security performance.
Hence the proof of Theorem \ref{thm:approximation} is concluded.

\section{A Lemma for Proof of Proposition \ref{prop:boundforR}}
\label{sec:appendix_MonotoneConvergence}

Here we prepare the following well-known fact in elementary analysis, which will be used in the proof of Proposition \ref{prop:boundforR} given in Appendix \ref{sec:proof_boundforR}:
\begin{lemma}
\label{lem:MonotoneConvergence}
Let $\{f_n\}_{n = 1}^{\infty}$ be a sequence of nonnegative continuous functions on the same open interval $I = (a,b)$, whose sum $\sum_{n = 1}^{\infty}f_n$ converges to a continuous function $f$ at every point in $I$.
If all of the improper integrals $\int_{a}^{b}f_n(x)\,dx$ and $\int_{a}^{b}f(x)\,dx$ exist and converge, then $\lim_{n \to \infty} \int_{a}^{b}\sum_{i = 1}^{n}f_i(x)\,dx = \int_{a}^{b}f(x)\,dx$.
\end{lemma}

From now, we give a proof of this result for the sake of completeness.
In the proof, we use the following two facts, which can be found in most of undergraduate textbooks of elementary analysis:
\begin{lemma}
[{Dini's Theorem; see e.g.\ \cite[p.151]{Car}}]
\label{lem:Dini'sTheorem}
Let $I = \left[a,b\right]$ be a closed interval, and let $\{g_i\}_{i=1}^{\infty}$ be an increasing sequence of continuous functions $g_i$ on $I$ which converges to another continuous function $g$ at every point in $I$; i.e.\ $g_{i-1}(x) \leq g_i(x) \to g(x)$ when $x \in I$.
Then the convergence is uniform; i.e.\ for any $\varepsilon > 0$, there is an index $n$ such that $|g(x) - g_i(x)| < \varepsilon$ for every $i \geq n$ and $x \in I$.
\end{lemma}
\begin{lemma}
[{see e.g.\ \cite[Theorem 10.5]{Car}}]
\label{lem:integral_uniform_convergence}
Let $\{g_i\}_{i=1}^{\infty}$ be a sequence of continuous functions on the same closed interval $I = \left[a,b\right]$ which converges uniformly to a function $g$ on $I$.
Then $\int_{a}^{b}g_i(x)\,dx$ converges to $\int_{a}^{b}g(x)\,dx$ when $i \to \infty$.
\end{lemma}

Now we start to prove Lemma \ref{lem:MonotoneConvergence}.
First, note that the function $f$ is nonnegative on $I$ by the assumption.
Then, given an arbitrary $\varepsilon > 0$, the assumption on convergence of the improper integral $\int_{a}^{b}f(x)\,dx$ infers that
\begin{equation}
\label{eq:MonotoneConvergence_1}
0 \leq \int_{a}^{b}f(x)\,dx - \int_{a'}^{b'}f(x)\,dx < \frac{\varepsilon}{\,2\,}
\end{equation}
for some $a < a' \leq b' < b$.
Now by the assumption, the sequence $\{g_i\}_{i=1}^{\infty}$ defined by $g_i = \sum_{n = 1}^{i}f_n$ is an increasing sequence of continuous functions on the closed interval $I' = \left[a',b'\right]$ which converges to the continuous function $f$ at every point in $I'$.
Thus Lemma \ref{lem:Dini'sTheorem} infers that the convergence is uniform; it follows that $\int_{a'}^{b'}g_i(x)\,dx \to \int_{a'}^{b'}f(x)\,dx$ when $i \to \infty$ by Lemma \ref{lem:integral_uniform_convergence}.
Therefore there is an index $n$ such that
\begin{equation}
\label{eq:MonotoneConvergence_2}
\int_{a'}^{b'}f(x)\,dx - \int_{a'}^{b'}g_i(x)\,dx < \frac{\varepsilon}{\,2\,} \mbox{ for any } i \geq n \enspace.
\end{equation}
By combining (\ref{eq:MonotoneConvergence_1}) and (\ref{eq:MonotoneConvergence_2}), we have
\setlength{\arraycolsep}{0.0em}
\begin{eqnarray}
0 &{}\leq{}& \int_{a}^{b}f(x)\,dx - \int_{a}^{b}g_i(x)\,dx \nonumber\\
&{}\leq{}& \int_{a}^{b}f(x)\,dx - \int_{a'}^{b'}g_i(x)\,dx 
< \frac{\varepsilon}{\,2\,} + \frac{\varepsilon}{\,2\,} = \varepsilon \nonumber
\end{eqnarray}
\setlength{\arraycolsep}{5pt}\noindent
for any $i \geq n$ (note that $0 \leq g_i \leq f$ for each $i$).
This means that $\int_{a}^{b}g_i(x)\,dx$ converges to $\int_{a}^{b}f(x)\,dx$ when $i \to \infty$, as desired.
Hence the proof of Lemma \ref{lem:MonotoneConvergence} is concluded.

\section{Proof of Proposition \ref{prop:boundforR}}
\label{sec:proof_boundforR}

Here we give the proof of Proposition \ref{prop:boundforR}.
First, let $\mathcal{Q}_n = (X_n,\omega_n)$ denote the Gauss-Legendre QS of order $n$ (see Section \ref{subsec:c-ind_optimal} for definition), therefore the set $X_n$ consists of $n$ zeroes of the Legendre polynomial $L_n(t)$.
In our proof of Proposition \ref{prop:boundforR}, we use the following result, which is directly derived from the latter part of Inequality (2.18) in \cite[Corollary 1]{FP1} by choosing the parameter $\lambda = 1/2$:
\begin{lemma}
[{\cite[Corollary 1]{FP1}}] 
\label{lem:Legendre_Bessel_2}
We have
\begin{displaymath}
\frac{ \omega_n(\xi) }{ \sqrt{ 1-\xi^2 } } \leq \frac{ \pi }{ n + 1/2 }
\end{displaymath}
for all $n \geq 1$ and all $\xi \in X_n$.
\end{lemma}

Now we start the proof of Proposition \ref{prop:boundforR}.
Recall that $c$ is now assumed to be even; put $c = 2n$.
First, by combining Proposition \ref{prop:c-ind_Requal} and Theorem \ref{thm:SRVandQS}, we have
\setlength{\arraycolsep}{0.0em}
\begin{eqnarray}
\mathcal{R}_{\ell,\mathcal{P}_c} &{}={}& E_p \left[ \sqrt{p(1-p)} \right] \nonumber\\
&{}={}& \sum_{\xi \in X_n} \sqrt{ \frac{ 1+\xi }{ 2 } \cdot \frac{ 1-\xi }{ 2 } } \cdot \frac{ \omega_n(\xi) }{ C_n\sqrt{ 1-\xi^2 } } \nonumber\\
&{}={}& \frac{1}{2C_n} \sum_{\xi \in X_n} \omega_n(\xi) = \frac{1}{C_n} \enspace ,\nonumber
\end{eqnarray}
\setlength{\arraycolsep}{5pt}\noindent
where $C_n = \sum_{\xi \in X_n} \omega_n(\xi) / \sqrt{ 1-\xi^2 }$.
Now Lemma \ref{lem:Legendre_Bessel_2} gives us that $C_n \leq n\pi/(n+1/2) = c\pi/(c+1)$, thus the former claim follows.

For the proof of the latter claim, it suffices to show that $\lim_{n \to \infty} C_n = \pi$.
Now it follows that
\begin{displaymath}
\frac{1}{\sqrt{1 - t^2}} = \sum_{i=0}^{\infty} \frac{1}{4^i} \binom{2i}{i} t^{2i} \geq \sum_{i=0}^{n-1} \frac{1}{4^i} \binom{2i}{i} t^{2i}
\end{displaymath}
for $-1 < t < 1$, therefore we have
\setlength{\arraycolsep}{0.0em}
\begin{eqnarray}
\frac{n\pi}{n + 1/2} \geq C_n
&{}\geq{}& \sum_{\xi \in X_n} \omega_n(\xi) \sum_{i=0}^{n-1} \frac{1}{4^i} \binom{2i}{i} \xi^{2i} \nonumber\\
&{}={}& \int_{-1}^{1} \sum_{i=0}^{n-1} \frac{1}{4^i} \binom{2i}{i} t^{2i}\,dt
\label{eq:limitofR_bound}
\end{eqnarray}
\setlength{\arraycolsep}{5pt}\noindent
(here (\ref{eq:limitofR_bound}) follows from the fact that the QS $\mathcal{Q}_n$ is of degree $2n-1$).
Now owing to Lemma \ref{lem:MonotoneConvergence} in Appendix \ref{sec:appendix_MonotoneConvergence} with $I = (-1,1)$ and $f_i(t) =4^{-i}\binom{2i}{i}t^{2i}$, the value (\ref{eq:limitofR_bound}) converges to $\int_{-1}^{1} \sum_i f_i(t)\,dt = \int_{-1}^{1} (1-t^2)^{-1/2}\,dt = \pi$ when $n \to \infty$, while $n\pi / (n + 1/2)$ also converges to $\pi$.
Thus we have $\lim_{n \to \infty} C_n = \pi$, as desired.
Hence the proof of Proposition \ref{prop:boundforR} is concluded.

\section{Proof of Proposition \ref{prop:limit_p0}}
\label{sec:proof_limit_p0}

Here we give the proof of Proposition \ref{prop:limit_p0}.
First, we define $\theta_n$ to be the unique value such that $0 < \theta_n < \pi$ and $-\cos \theta_n$ is the smallest zero of the Legendre polynomial $L_n(t)$ (recall that the zeroes of $L_n(t)$ lie in the open interval $(-1,1)$).
Then we have the following result, which will be used in the proof of Proposition \ref{prop:limit_p0}:
\begin{lemma}
[{\cite[p.264]{Gat}}]
\label{lem:Legendre_Bessel_1}
For any $n$, we have
\begin{displaymath}
\frac{ j_1 }{ \sqrt{ (n + 1/2)^2 + a_2 } } 
< \theta_n 
< \frac{ j_1 }{ \sqrt{ (n + 1/2)^2 + a_1 } } \enspace ,
\end{displaymath}
where $a_1 = 1/12$ and $a_2 = 1/4 - 1/\pi^2$.
\end{lemma}

From now, we prove Proposition \ref{prop:limit_p0}.
First, by the definitions of $p_0$ and $\theta_n$, Theorem \ref{thm:SRVandQS} infers that
\begin{displaymath}
p_0 = \frac{ 1 - \cos\theta_n }{ 2 }
= \sin^2 \frac{\theta_n}{2} \enspace ,
\end{displaymath}
therefore $\sigma(p_0) = 1/\tan(\theta_n/2)$.
Thus by Lemma \ref{lem:Legendre_Bessel_1}, we have
\setlength{\arraycolsep}{0.0em}
\begin{eqnarray}
\frac{ 1 }{ \tan( j_1 / \sqrt{ (c+1)^2 + a'_1 } ) }
&{}<{}& \sigma(p_0)\label{eq:bound_p0_1}\\
\hspace*{4em}&{}&\hspace*{-4em}< \frac{ 1 }{ \tan( j_1 / \sqrt{ (c+1)^2 + a'_2 } ) }  \enspace ,\label{eq:bound_p0_2}
\end{eqnarray}
\setlength{\arraycolsep}{5pt}\noindent
where $a'_i = 4a_i$ (namely $a'_1 = 1/3$ and $a'_2 = 1 - 4\pi^{-2}$).
Moreover, an elementary analysis gives us
\setlength{\arraycolsep}{0.0em}
\begin{eqnarray}
&&\frac{ 1 }{ c\tan( j_1 / \sqrt{ (c+1)^2 + a'_i } ) } \nonumber\\
&&= \frac{ \cos( j_1 / \sqrt{ (c+1)^2 + a'_i } ) }{ j_1 / \sqrt{ (1+1/c)^2 + a'_i/c^2 } } \cdot \frac{ j_1 / \sqrt{ (c+1)^2 + a'_i } }{ \sin( j_1 / \sqrt{ (c+1)^2 + a'_i } ) } \nonumber
\end{eqnarray}
\setlength{\arraycolsep}{5pt}\noindent
for $i = 1,2$, which converges to $1/j_1$ when $c \to \infty$.
Hence the claim of Proposition \ref{prop:limit_p0} follows, concluding the proof.

\section{Proof of Theorem \ref{thm:boundoflength_general}}
\label{sec:proof_boundoflength_general}

Here we give the proof of Theorem \ref{thm:boundoflength_general}.
First, we have $\mathcal{R} = \mathcal{R}_{\ell,\mathcal{P}} \to 1/\pi$ by Proposition \ref{prop:boundforR} (see also Proposition \ref{prop:c-ind_Requal}) and $c^{-1} \sigma(p_0) \to 1/j_1$ by Proposition \ref{prop:limit_p0} when $c \to \infty$.
Thus the parameter $x_1$, which is determined by $x_1 r(x_1) = \eta_1 \mathcal{R} \sigma(p_0) / c$, converges when $c \to infty$ to $x_\infty$ given in the statement (note that the continuous function $t r(t)$ is strictly increasing for $t>0$ and its image is the whole infinite interval $(0,+\infty)$, therefore such $x_\infty$ is uniquely determined).
Similarly, the other parameter $x_2$ converges to $\infty$ when $c \to \infty$, since $x_2 r(x_2) = \eta_2 \mathcal{R} \sigma(p_0) / c \to \infty$.
Now by (\ref{eq:newformula_length}), we have
\setlength{\arraycolsep}{0.0em}
\begin{eqnarray}
&&\frac{ m }{ c^2 \log(1/\varepsilon_1) } \nonumber\\
&&= \frac{ 1 }{ \mathcal{R} (1 - \eta_1 - \eta_2 c^{-1} ) } \cdot \frac{ \sigma(p_0) }{ c }\!\left( \frac{ 1 }{ x_1 } + \frac{ \log(1/\varepsilon_2) }{ x_2 \log(1/\varepsilon_1) } \right)\! \enspace ,\nonumber
\end{eqnarray}
\setlength{\arraycolsep}{5pt}\noindent
while the above argument shows that
\begin{displaymath}
\mathcal{R} (1 - \eta_1 - \eta_2 c^{-1} ) \to \frac{ 1 - \eta - \eta' }{ \pi }
\mbox{ and } \frac{ \sigma(p_0) }{ c } \to \frac{1}{j_1}
\end{displaymath}
when $c \to \infty$.
Moreover, by the assumption on value $\log(1/\varepsilon_2) / \log(1/\varepsilon_1)$ we have that
\begin{displaymath}
\frac{ \log(1/\varepsilon_2) }{ x_2 \log(1/\varepsilon_1) } \to 0 \mbox{ when } c \to \infty \enspace .
\end{displaymath}
Thus we have
\begin{displaymath}
\frac{ m }{ c^2 \log(1/\varepsilon_1) } \to \frac{ \pi }{ (1-\eta-\eta') j_1 x_\infty } \enspace ,
\end{displaymath}
therefore the claim follows.
Hence the proof of Theorem \ref{thm:boundoflength_general} is concluded.
\end{document}